\documentclass{elsart}

\usepackage{graphicx}
\graphicspath{{./images/}}
\usepackage{cite}
\usepackage{amsmath}
\usepackage{amssymb}
\usepackage[latin1]{inputenc}

\renewcommand{\L}{\mathrm{L}}
\newcommand{\R}{\mathrm{R}}
\newcommand{\C}{\mathrm{C}}
\newcommand{\eff}{\mathrm{eff}}
\newcommand{\ac}{\mathrm{ac}}
\newcommand{\TM}{\mathbf{T}}
\renewcommand{\d}{\mathrm{d}}
\newcommand{\I}{\ensuremath{\mathrm{i}}}
\newcommand{\T}{\ensuremath{\mathcal{T}}}

\begin{document}

\begin{frontmatter}

\title{Transport suppression in heterostructures driven by an ac gate voltage}
\author[a]{Miguel Rey},
\author[b]{Michael Strass},
\author[b]{Sigmund Kohler},
\author[c]{Fernando Sols}, and
\author[b]{Peter H\"anggi}
\address[a]{Departamento de F{\chardef\i="10 \'\i}sica Te\'orica de la Materia
  Condensada, Universidad Aut\'onoma de Madrid, E-28049 Madrid, Spain}
\address[b]{Institut f\"ur Physik, Universit\"at Augsburg,
Universit\"atsstra\ss e~1, D-86135 Augsburg, Germany}
\address[c]{Departamento de F{\chardef\i="10 \'\i}sica de Materiales,
  Universidad Complutense de Madrid, E-28040, Spain}

\begin{abstract}
We explore the possibility of inducing in heterostructures driven by an ac gate
voltage the coherent current suppression recently found for nanoscale
conductors in oscillating fields. The destruction of current is fairly independent of the
transport voltage, but can be controlled by the driving amplitude and
frequency. Within a tight-binding approximation, we obtain
analytical results for the average current in the presence of
driving. These results are compared against an exact numerical
treatment based on a transfer-matrix approach.
\end{abstract}

\begin{keyword}
quantum transport \sep driven systems \sep heterostructures \sep
tunnelling \PACS
05.60.Gg \sep 
85.65.+h \sep 
72.40.+w  
\end{keyword}
\end{frontmatter}

\section{Introduction and modelling}
\label{sec:intro}

The study of electron transfer comprises a rich variety of systems in
many different areas such as chemistry, biology, and life
sciences~\cite{Kuznetsov1995a, Bixon1999a}. Although electron transfer
processes are mainly attributed to electrochemical applications, they are
conceptually related to molecular
electronics~\cite{Nitzan2001a,Hanggi2002elsevier,Cuniberti2005a} and
electron transport in low dimensional
materials in solid-state physics.  In that context,
semiconductor heterostructures represent a popular physical system
for the investigation of mesoscopic
transport~\cite{Beenakker1991a, Imry1986a, Blanter2000a} and tunnelling
phenomena~\cite{Esaki1970a, Tsu1973a, Chang1974a, Sollner1983a,
Capasso1990a}. The main reason for this is the high mobility and
the rather long mean free path of the charge carriers populating
them. Standard beam epitaxy techniques make the accurate
growth of alloys of such materials on substrates possible, and the nearly
identical lattice parameters, together with the possibility of
controlling the band gap, turns the combination
GaAs/Al$_x$Ga$_{1-x}$As into an ideal candidate for building
complex low dimensional structures with quantum wells and tunnel
barriers.
Moreover, these setups open various ways to study tunnelling in
time-dependent systems \cite{Grifoni1998a, Platero2004a,
Kohler2005a}.  A straightforward possibility for introducing a
time-dependence is the application of an ac transport voltage
which only modulates the energies of the electrons in the leads
while the potential inside the mesoscopic region remains
time-independent.  This kind of driving allows for a description
within Tien-Gordon theory \cite{Tien1963a} which expresses the dc
current in terms of the static transmission and an effective
distribution function for the lead electrons. If the
time-dependence enters via an external microwave field or an ac
gate voltage, however, such an approach is generally insufficient~\cite{Camalet2004a}.

A remarkable difference with respect to the static situation is
the emergence of inelastic transport channels stemming from the
emission or absorption of quanta of the driving field. For a
periodically time-dependent transport situation, however, we
expect the transmission probabilities and, consequently, the
resulting current to be time-dependent as well. This follows
indeed from a recently presented Floquet theory for the transport
through driven tight-binding systems \cite{Camalet2004a,
Kohler2005a}. For the computation of the dc current, this approach
justifies the applicability of a Landauer-like current formula
where the static transmission is replaced by the time-averaged
transmission of the time-dependent system.

The transmission of both the elastic and the inelastic transport
channels can depend sensitively on the driving parameters;
the contribution of certain channels can even vanish.
For the transport across two barriers which enclose an oscillating
potential well, Wagner \cite{Wagner1994a} showed that it is
possible to suppress the contribution of individual inelastic
scattering channels. The total current, however, is given by the
sum over all channels, and thus it is not possible to isolate the
contribution of a single channel in a current measurement.
By contrast, in the case of transport through a two-level system
with attached leads, it has been found that driving with a dipole
field has directly observable consequences. There, the driving not
only affects the contribution of individual transport channels,
but the dc current can be suppressed almost entirely
\cite{Lehmann2003a, Kohler2004a}. Therefore, for the appearance of
this \textit{coherent current suppression}, it is essential that
the central region consists of at least two weakly coupled wells
which oscillate relative to each other \cite{Camalet2004a}.

\begin{figure}[t]
  \centering
  \includegraphics[width=7.5cm]{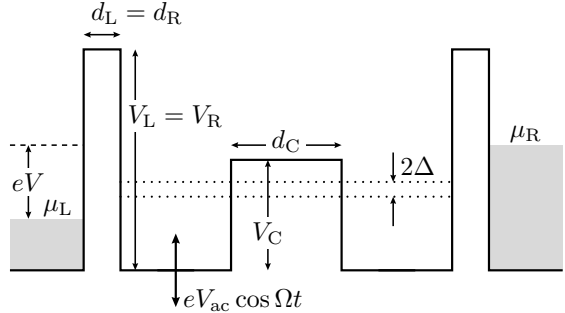}
  \caption{Model potential for the double-well heterostructure.
  In the numerical calculations, we employ barriers with the heights
  $V_{\L}=V_{\R}= 90\,\mathrm{meV}$, $V_{\C}=40\,\mathrm{meV}$ and the widths
  $d_{\L}=d_{\R}=5\,\mathrm{nm}$, $d_{\C}=15\,\mathrm{nm}$.
  The dotted lines mark the energy of a metastable tunnel doublet with
  splitting energy $2\Delta$. The left well is subject to an electric
  dipole field generated by an alternating gate voltage with amplitude
  $V_{\ac}$.}
  \label{fig:hetero-model}
\end{figure}%
In this work, we explore the possibility of coherent current
suppression in double-well heterostructures. Thereby, we compare two
theoretical approaches to describe coherent transport in quantum-well
structures: The transfer-matrix method and a tight-binding approach. As a model we
consider the triple-barrier structure sketched in
Fig.~\ref{fig:hetero-model} where the driving enters via an
oscillating gate voltage which modulates the bottom of the left
well. The applied transport voltage is assumed to shift the Fermi
energy of the left lead by $-eV$ with $-e$ being the electron
charge.  We note that since the time-dependent gate voltage affects
only one well, the structure depicted in Fig.~\ref{fig:hetero-model}
is sufficiently asymmetric to also act as an electron pump,
i.e.\ to induce a non-zero current for $eV=0$ \cite{Kohler2005a}.
In this work, however, we focus on the transport properties in the
presence of a finite bias voltage.

For the exact numerical computation of the transmission
probabilities, we employ the transfer-matrix method developed by
Wagner \cite{Wagner1995a}, which is reviewed in
Section~\ref{sec:transfer-matrix}. In Section~\ref{sec:tight-binding},
we introduce the related tight-binding system for which the
transport properties can be calculated analytically within a
high-frequency approximation scheme \cite{Kohler2004a}. The
predictions from the perturbative approach are compared to the
exact solution in Section~\ref{sec:results}.

\section{Transfer-matrix method}
\label{sec:transfer-matrix}

Following Landauer \cite{Landauer1957a}, we consider the coherent
mesoscopic transport as a quantum mechanical scattering process. The
central idea of this approach is the assumption that sufficiently
far from the scattering region, the electronic single-particle
states are plane waves and that their occupation probability is
given by the Fermi function with the chemical potential depending on
the applied voltage. The unitarity of evolution under coherent ac
driving allows us to write the resulting currents as
\cite{Wagner1999a}
\begin{equation}
  \label{eq:Landauer}
  I = \frac{e}{h} \int \d E 
  \,[T_{\R\L}(E) f_{\L}(E) - T_{\L\R}(E) f_{\R}(E) ] ,
\end{equation}
where $T_{\R\L}(E)$ denotes the total transmission probability
--- i.e. summed over transverse modes and outgoing inelastic channels --- of
an electron with energy $E$ from the left lead to the right lead
while $T_{\L\R}(E)$ describes the respective scattering from the
right to the left lead. For time-independent conductors, the
time-reversal symmetry of the quantum mechanical scattering
process together with the energy conservation ensures $T_{\R\L}(E)
= T_{\L\R}(E)$ such that, in the absence of a transport voltage,
the current vanishes. This is not the case for a general
time-dependent structure \cite{Kohler2005a}.

When the total Hamiltonian is time-periodic due to an external
driving field, $H(x,t)=H(x,t+\T)$, one can apply Floquet theory
\cite{Sambe1973a, Shirley1965a, Grifoni1998a}. It states that the
corresponding time-dependent Schr\"odinger equation has a complete
set of solutions of the form
\begin{equation}
  \label{eq:Fstates}
  \psi_{\alpha}(x,t)=\exp(-\I \epsilon_{\alpha}t/\hbar)
  u_{\alpha}(x,t) ,
\end{equation}
where $u_{\alpha}(x,t)=u_{\alpha}(x,t+\T)$ denotes the so-called
Floquet states, and $\epsilon_{\alpha}$ the so-called quasienergies
in analogy to the quasimomenta of Bloch theory.

Owing to their time-periodicity, we can decompose the Floquet
states into a Fourier series
\begin{equation}
  \label{eq:Fstates-Fourier}
  u_{\alpha}(x,t) = \sum_{n=-\infty}^\infty \mathrm{e}^{-\I n \Omega t}\,u_{\alpha,n}(x).
\end{equation}
The form \eqref{eq:Fstates-Fourier} of the Floquet states suggests
that during the scattering, an electron with initial energy $E$
evolves into a \textit{coherent} superposition of states with
energies $E' = E+n\hbar\Omega$. The arbitrary integer $n$ is
referred to as the sideband index; the inelastic channels are
called sidebands. We emphasise that, despite the existence of a
band bottom,  the summation over $n$ in
Eq.~\eqref{eq:Fstates-Fourier} is unrestricted.

A proper calculation of the dc current through a time-dependent
scatterer must now include these inelastic channels, and the
Landauer formula of Eq.~\eqref{eq:Landauer} has to be conveniently
generalised to take them into account. Since the Floquet
scattering states can be thought of as having been created from
the orthogonal dc states by  adiabatically switching on the
driving, they too must be orthogonal, so that it is sufficient to
sum over channels incident from both leads \cite{Wagner1999a}.

In the transfer-matrix method described below, the Floquet states
are decomposed into plane waves throughout the driven
heterostructure, so that they can be appropriately matched with
the scattering channels in the leads. This decomposition allows us
to separate the time- and the space-dependent parts of the wave
function, thus obtaining directly the time-independent
probabilities that go into Eq.~\eqref{eq:Landauer}.

For a spatially constant potential with a time-dependent gate
voltage $V_\ac(t)$, the solution of the Schr\"odinger equation is
readily obtained to read
\begin{equation}
  \psi (E,z,t) = \sum^{+\infty}_{n=-\infty} \psi_n(z)
  \exp \left\{-\frac{\I}{\hbar} (E+n\hbar \Omega)t -\I\phi(t)
  \right\}
  \label{eq:Schr-sol}
\end{equation}
with the accumulated phase
\begin{equation}
\phi(t) = \frac{e}{\hbar} \int^{t}_{0} \d t' V_{\ac}(t') = \phi(t+\T) .
\label{eq:phase}
\end{equation}
Its time-periodicity follows from the zero time-average of the gate
voltage.

Neighbouring layers of a heterostructure may have different ac
voltages applied in addition to different band-edges. As a
consequence, the wave functions in Eq.~\eqref{eq:Schr-sol} which
solve the Schr\"odinger equation in each layer do not coincide in
the general case. The solution for the complete system has to be
constructed by matching the corresponding wave functions at the
interfaces between layers. With this goal in mind, we assume that
the wave function~\eqref{eq:Schr-sol} is a solution of the
Schr\"odinger equation for the time-dependent Hamiltonian
\begin{eqnarray}
    \label{eq:H(t)-hetero}
    H(z,t) &=& H_0(z) + e V_{\ac}(t) \nonumber \\
           &=& -\frac{\hbar^2}{2} \frac{\partial}{\partial z}
           \frac{1}{m(z)} \frac{\partial}{\partial z} + V(z)
           + e V_{\ac}\cos \Omega t,
\end{eqnarray}
and that, moreover, $\psi_n(z)$ is an eigenfunction of the
time-independent Hamiltonian $H_0$ with the spatially piecewise
constant effective mass $m(z)$, and has the general form
\begin{equation}
    \label{eq:statsol}
    \psi_n (z) = A_n \exp (k_n z) + B_n \exp (-k_n z).
\end{equation}
The wave vector
\begin{equation}
    k_n = \left[2m(V-E-n\hbar \Omega)\right]^{1/2}
\end{equation}
describes travelling as well as decaying waves (bound states) for
complex and real values of $k_n$, respectively. The matching
conditions at an interface follow from the fact that both the wave
function and the flux have to be continuous, i.e.\ at $z=z_0$
\begin{eqnarray}
    \lim_{z\rightarrow z_0^+} \psi (z,t) &=& \lim_{z\rightarrow z_0^-} \psi (z,t) \nonumber \\
    \lim_{z\rightarrow z_0^+} \frac{1}{m(z)}\frac{\partial}{\partial z} \psi (z,t) &=&
    \lim_{z\rightarrow z_0^-} \frac{1}{m(z)}\frac{\partial}{\partial z} \psi (z,t).
    \label{eq:match-cond}
\end{eqnarray}
This yields an infinite system of algebraic equations for the
coefficients $A_n$ and $B_n$ in each layer. Inserting the Fourier
expansion of the phase in Eq.~\eqref{eq:phase},
\begin{equation}
    \exp \left\{-\frac{\I e}{\hbar\Omega}V_{\ac} \sin \Omega t \right\} =
    \sum_{n'= -\infty}^{+\infty} J_{n'} \left(\frac{e V_{\ac}}{\hbar
    \Omega}\right)\exp (-\I n' \Omega t), \label{eq:Bessel.exp}
\end{equation}
where $J_{n'}$ is the $n'$-th order Bessel function of the first
kind, allows one to recast these equations for an interface
between layers I and II at $z=z_i$ in matrix form:
\begin{equation}
    \label{eq:TMs(1)}
    \TM^{\mathrm{I}}_{z_i}
    \left( \begin{array}{c} A^{\mathrm{I}}_n \\ B^{\mathrm{I}}_n \end{array} \right) =
    \TM^{\mathrm{II}}_{z_i}
    \left( \begin{array}{c} A^{\mathrm{II}}_n \\ B^{\mathrm{II}}_n \end{array} \right).
\end{equation}
The matrices $\TM^{\mathrm{I}}_{z_i}$ and
$\TM^{\mathrm{II}}_{z_i}$ with elements
$T^{\mathrm{I}}_{z_i;n,n'}$ and $T^{\mathrm{II}}_{z_i;n,n'}$,
respectively, are of infinite dimension, and contain the
coefficients of all possible scattering channels, i.e. photon
exchanges between the incoming electron and the driving field, at
either side of the interface. Their precise form depends on
whether or not a layer is affected by the time-dependent gate
voltage. The transfer matrix $\TM_{z_i \rightarrow z_j}$ between
two sides of a layer of width $z_j-z_i$ is then defined as
\cite{Wagner1993a}
\begin{equation}
    \TM_{z_i \rightarrow z_j} = \TM^{\mathrm{I}}_{z_j} \big(
    \TM^{\mathrm{I}}_{z_i} \big)^{-1}. \label{eq:TM-def}
\end{equation}
This definition in terms of {layers}, rather than using a
similar one across {interfaces}, is physically more
sensible, since the matrices depend on the properties of just one
layer. They are also numerically easier to implement, as there are
fewer qualitatively different matrices to deal with. To calculate
the total transfer matrix across the structure, we have to
multiply the matrices across all different layers and obtain
\begin{equation}
    \TM_{\L \rightarrow \R} = \TM_{z_{\R}} \TM_{z_j \rightarrow z_{\R}} \cdots
    \TM_{z_{\L} \rightarrow z_i} \TM_{z_{\L}},
    \label{eq:TM-totalTM}
\end{equation}
where $\TM_{z_{\L}}$ and $\TM_{z_{\R}}$ represent the initial and
final matrices at the ends of the heterostructure. With the elements
$T^{n,n'}_{\L \rightarrow \R}$ of the total transfer matrix we can find
the probability that an electron with energy $E+n \hbar \Omega$ in
lead $\L$ is scattered into a channel with energy $E' = E+n' \hbar
\Omega$ in lead $\R$, with integer $n,n'$. The diagonal elements
$T^{n,n}_{\L \rightarrow \R}$ are closely related to the (static)
transmission probability $T_{\R \L}$, while the off-diagonal
elements $T^{n\neq n'}_{\L \rightarrow \R}$ describe the effects
of the absorption or emission of $n-n'$ photons on the
transmission probability of the electron. For flat conduction
bands on both sides of the heterostructure, the wave functions in
the contacts are plane waves, and in this case the proper boundary
conditions to describe an electron incident from, say, the
left-hand side at energy $E$ are $A^n_{\L}=\delta_{n,0}$ and
$B^n_{\R}=0$. The transmission probability in sideband $n$ is then
defined as
\begin{equation}
    T^n_{\R \L} =
    \frac{k^n_{\R}}{k^0_{\L}}\frac{m_{\L}}{m_{\R}}\left|\frac{A^n_{\R}}{A^0_{\L}}\right|^2,
\label{eq:TM-trans}
\end{equation}
where $k^n_{\R}$ and $k^0_{\L}$ represent the wave vectors on the
right- and left-hand side in sidebands $n$ and 0, respectively.

In a numerical implementation of the transfer-matrix technique, it is
necessary to truncate the infinite matrices.  Thereby for consistency,
a proper cut-off has to be so large that unitarity of the scattering
process is preserved, i.e.
\begin{equation}
    \sum_{n= -\infty}^{+\infty} T^n_{\R \L} + \sum_{n=
    -\infty}^{+\infty} R^n_{\L \L} = 1,
    \label{eq:TM-unitarity}
\end{equation}
where $R^n_{\L \L}$ represents the reflection probability of an
electron to be reflected from energy $E$ into a sideband at energy
$E+n\hbar \Omega$ on the same side. The number of sidebands that
need to be taken into account to meet a given accuracy (which in
our calculations was set to $10^{-17}$) depends essentially on the
ratio $e V_{\ac}/\hbar \Omega$, as this is the argument of the
Bessel function $J_n$ that determines the weight of each sideband
$n$. To proceed, one starts at the initial value of $V_{\ac}=0$
with a tentative number of sidebands, and increases it for growing
driving amplitudes if a check with Eq.~\eqref{eq:TM-unitarity}
suggests that unitarity is breaking down. When particle number
conservation is restored one can go to a higher $V_{\ac}$.
Transfer matrices such as those employed here have the advantage
of being easily scalable to arbitrarily complex structures. The
combination of flexibility in structural properties and numerical
accuracy makes this method well-suited to the study of strongly
driven semiconductor heterostructures.

\section{Tight-binding approximation}
\label{sec:tight-binding}

\begin{figure}[t]
  \centering
  \includegraphics[width=7.5cm]{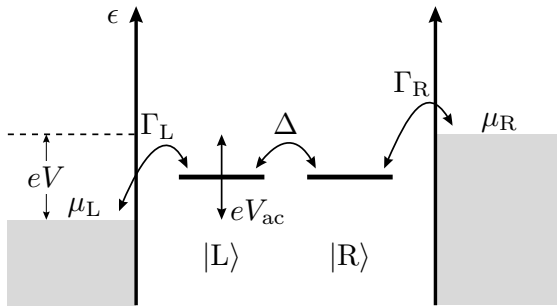}
  \caption{Schematic energy diagram of a symmetric double-well structure
    in the tight-binding approximation. The tunnelling matrix element is
    given by $\Delta$ and each well couples with $\Gamma_{\L}$ and
    $\Gamma_{\R}$ to the
    associated lead. The state $|\L\rangle$ in the left well is driven by
    a oscillating gate voltage with amplitude $V_{\ac}$. In
    addition, an external bias $V=(\mu_{\R}-\mu_{\L})/e$ is applied.}
  \label{fig:tb-model}
\end{figure}%
A different approach to study resonant tunnelling in a driven
double-well structure is based on the adoption of a tight-binding
(TB) picture. Figure~\ref{fig:tb-model} depicts the TB configuration
corresponding to the heterostructure introduced above.

\subsection{The model}
\label{sec:tb-model}

Then, the Hamiltonian of the system is given by
\begin{equation}
  \label{eq:H(t)}
  H(t) = H_{\mathrm{wells}}(t) + H_{\mathrm{leads}} +
  H_{\mathrm{contacts}}\, ,
\end{equation}
where the individual terms describe the driven quantum wells, the
electron reservoirs in the leads, and the coupling of the left and the
right well to the respective neighbouring lead. For
simplicity, we neglect the electron spin.

Within the framework of the TB approximation, the time-dependent
quantum-well Hamiltonian reads
\begin{equation}
  \label{eq:Hwells(t)}
  H_{\mathrm{wells}}(t) =
  \sum_{\ell,\ell'} H_{\ell,\ell'}(t) c_{\ell}^{\dag} c_{\ell'} =
  -\Delta(c_{\L}^{\dag} c_{\R}  + c_{\R}^{\dag} c_{\L})
  + e V_{\ac} \cos(\Omega t) c_{\L}^{\dag} c_{\L}\, .
\end{equation}
An electron can be localised in the left or right well, whereupon
the fermion operator $c_{\ell}$ ($c_{\ell}^{\dagger}$) annihilates
(creates) an electron in the respective well ($\ell=\L,\R$). These
localised states $|\L\rangle$ and $|\R\rangle$ are coupled by the
tunnelling matrix element $\Delta$. For convenience, the energy
scale is set such that the on-site energies of the two resonant TB
levels are zero and lie exactly halfway in between the transport
bias window defined by the chemical potentials $\mu_{\L}$ and
$\mu_{\R}$. The second term of the Hamiltonian \eqref{eq:Hwells(t)}
accounts for the harmonic driving of the traversing electrons in the
left well via an oscillating gate voltage with amplitude $V_{\ac}$
and period $\T=2\pi/\Omega$.

The leads are modelled as ideal Fermi gases with the Hamiltonian
\begin{equation}
  \label{eq:Hleads}
  H_{\mathrm{leads}} = \sum_{\ell, q} \epsilon_{\ell q} c_{\ell
  q}^{\dag} c_{\ell q},
\end{equation}
where $c_{\ell q}$ ($c_{\ell q}^{\dag}$) annihilates (creates) an
electron in the lead with energy $\epsilon_{\ell q}$ with
$\ell=\L,\R$. As an initial condition, we employ the
grand-canonical ensembles of electrons in the leads at inverse temperature
$\beta=1/k_B T$. Therefore, the lead electrons are characterised by the
equilibrium Fermi distribution $f_{\ell}(\epsilon_{\ell
q})=\{1+\exp[-\beta(\epsilon_{\ell q}-\mu_{\ell})]\}^{-1}$.

The localised state in each well couples via the tunnelling matrix
element $V_{\ell q}$ to the state $|\ell q\rangle$ in the
respective lead. The Hamiltonian which describes this interaction
has the form
\begin{equation}
  \label{eq:Hcontacts}
  H_{\mathrm{contacts}} = \sum_{\ell,q} V_{\ell q} c^{\dag}_{\ell
  q} c_{\ell} + \mathrm{H.c.}
\end{equation}
The lead--well coupling is entirely specified by the spectral density
$\Gamma_{\ell}(\epsilon)=2\pi\sum_{q} |V_{\ell q}|^2
\delta(\epsilon-\epsilon_{\ell q})$. Since, for the system at hand,
the bandwidth of the conduction band of the leads is much larger
than the energy regime where transport happens, the spectral
densities are practically constant, i.e.
$\Gamma_{\ell}(\epsilon)=\Gamma_{\ell}$, which defines the so-called
wide-band limit.

\subsection{Floquet transport theory}
\label{sec:floquet-transport}

In order to determine the time-averaged dc current which matches
Eq.~\eqref{eq:Landauer} for the TB approximation, we employ a
generalised Floquet approach to solve the corresponding Heisenberg
equations of motion and derive an expression for the retarded
Green's function in terms of Floquet states. This result is applied
to study transport by evaluating the operator $I_{\ell}(t)=(\I e/\hbar)
[H(t),\sum_{q}c_{\ell q}^{\dag}c_{\ell q}]$ for the time-dependent
current through contact $\ell$.  In the next subsection we will also
derive a analytic expression for the current valid in the high
frequency limit with respect to the driving.

We start out by stating the solution of the Heisenberg equations
of motion for the lead operators, which is given for the left lead
by \cite{Camalet2004a}
\begin{equation}
  \label{eq:c_lead(t)}
  c_{\L q}(t)=c_{\L q}(t_0)\mathrm{e}^{-\I\epsilon_{\L q}(t-t_0)/\hbar}
  -\frac{\I V_{\L q}}{\hbar}\!\!\int\limits_{t_0}^{t}\! \d t'\,
  \mathrm{e}^{-\I\epsilon_{Lq}(t-t')/\hbar} c_{\L}(t').
\end{equation}
For the corresponding solution of $c_{\R q}(t)$, $\L$ has to be
substituted by $\R$. Now inserting
these solutions into the Heisenberg equations of motion for the
quantum wells yields the two coupled linear differential equations
\begin{align}
  \label{eq:cdot_L}
  \dot c_{\L} =& -\frac{\I e}{\hbar}V_{\ac}\cos(\Omega t)c_{\L}
  - \frac{\Gamma_{\L}}{2\hbar}c_{\L} + \frac{\I}{\hbar}\Delta
  c_{\R} + \xi_{\L}(t)\\
  \label{eq:cdot_R}
  \dot c_{\R} =& \frac{\I}{\hbar}\Delta c_{\L}
  -\frac{\Gamma_{\R}}{2\hbar}c_{\R} + \xi_{\R}(t).
\end{align}
Here, within the wide-band limit, the coupling to the leads has been
eliminated in favour of the spectral density $\Gamma_{\ell}$
and the fermionic fluctuation operator
\begin{equation}
  \label{eq:xi(t)}
  \xi_{\ell}(t)=-\frac{\I}{\hbar}\sum_q V^{*}_{\ell q}\,
  \mathrm{e}^{-\I\epsilon_{\ell q}(t-t_0)/\hbar}\, c_{\ell q}(t_0) .
\end{equation}
Assuming that the initial conditions are those of
the grand canonical ensemble, this Gaussian noise operator
satisfies
\begin{align}
  \label{eq:<xi(t)>}
  \langle\xi_{\ell}(t)\rangle &= 0,\\
  \label{eq:<xi(t)xi(t)>}
  \langle\xi^{\dag}_{\ell}(t)\,\xi_{\ell'}(t')\rangle
  &= \delta_{\ell\ell'}\frac{\Gamma_{\ell}}{2\pi\hbar^2}
  \int\!\! \d \epsilon\,
  \mathrm{e}^{\I\epsilon(t-t')/\hbar}f_{\ell}(\epsilon).
\end{align}
Then the operator for the time-dependent current through the left lead becomes
\cite{Camalet2004a}
\begin{equation}
  \label{eq:I(t)}
  I_{\L}(t) = \frac{e\Gamma_{\L}}{\hbar}c_{\L}^\dag(t)c_{\L}(t)
  -e\big[c_{\L}^\dagger(t)\xi_{\L}(t)+\xi_{\L}^\dag(t)c_{\L}(t)\big]
\end{equation}
with a corresponding expression for $I_{\R}(t)$. Here we made use of
the wide-band limit. To evaluate the time-dependent current, we thus
have to find the solution for the inhomogeneous set of the quantum
Langevin Eqs.~\eqref{eq:cdot_L} and \eqref{eq:cdot_R} of the
quantum-well operators, which is formally given by
\begin{equation}
  \label{eq:c(t)}
  c_{\ell}(t) = \int_{0}^{\infty}\!\!\d \tau\,
  \big[ G_{\ell \L}(t,t-\tau) \xi_{\L}(t-\tau) +
  G_{\ell \R}(t,t-\tau) \xi_{\R}(t-\tau) \big]
\end{equation}
in the stationary limit $t_{0}\rightarrow\infty$ with $\ell=\L,\R$.
What remains is to determine the retarded Green's function
$G(t,t-\tau)$. This is where Floquet theory comes into play by
making use of the $\T$-periodicity of the driving. Solving the
Floquet eigenvalue equation
\begin{equation}
  \label{eq:Feq-non-hermit}
  \Big( \mathcal{H}_{\mathrm{wells}}(t) - \I\Sigma
  - \I\hbar\frac{\partial}{\partial t} \Big) |u_{\alpha}(t)\rangle =
  (\epsilon_{\alpha}-\I\hbar\gamma_{\alpha}) |u_{\alpha}(t)\rangle
\end{equation}
of the physical problem at hand, where $\mathcal{H}(t) =
\sum_{\ell,\ell'}|\ell\rangle H_{\ell,\ell'}(t)\langle\ell'|$, we
get the Floquet states $|u_{\alpha}(t)\rangle$ and the
complex-valued quasienergies
$\epsilon_{\alpha}-\I\hbar\gamma_{\alpha}$. Note that the prior
equation is, in contrast to the usual Floquet equation,
non-Hermitian. This is due to the presence of the self-energy
$2\Sigma=|\L\rangle\Gamma_{\L}\langle\L| +
|\R\rangle\Gamma_{\R}\langle\R|$, which results from tracing out the
leads~\cite{Datta1995a}. Therefore, Eq.~\eqref{eq:Feq-non-hermit}
has to be solved also for its adjoint eigenstates
$|u_{\alpha}^{+}(t)\rangle$~\cite{Kohler2005a}. With the
corresponding expression for the propagator $U(t,t-\tau)$, the
retarded Green's function assumes the form
\begin{equation}
  \label{eq:G_retarded(t)}
  G(t,t-\tau) = \sum_{\alpha}
  \mathrm{e}^{-\I(\epsilon_{\alpha}/\hbar-\I\gamma_{\alpha})\tau}
  |u_{\alpha}(t)\rangle \langle u_{\alpha}^{+}(t-\tau)| \Theta(\tau),
\end{equation}
where $\Theta(\tau)$ is the Heaviside step function.

The dc current is now obtained by calculating the expectation value
$\langle I_{\L}(t)\rangle$ and averaging over one driving period.
This time-averaging will cancel those terms of $\langle
I_{\L}(t)\rangle$, which are responsible for a \T-periodic charging
of the wire. After eliminating backscattering
terms~\cite{Camalet2004a}, we arrive at the very compact form for
the final result
\begin{equation}
  \label{eq:Ibar}
  \bar I = \frac{e}{h} \sum_{n}\int\!\!\d \epsilon\,
  \big[ T^{(n)}_{\L\R}(\epsilon) f_{\R}(\epsilon)
  - T^{(n)}_{\R\L}(\epsilon) f_{\L}(\epsilon) \big],
\end{equation}
where the total transmission probabilities are given by
$T^{(n)}_{\L\R}(\epsilon)=\Gamma_{\L}\Gamma_{\R}|G_{\L\R}^{(n)}(\epsilon)|^{2}$
and
$T^{(n)}_{\R\L}(\epsilon)=\Gamma_{\L}\Gamma_{\R}|G_{\R\L}^{(n)}(\epsilon)|^{2}$,
which resemble the Fisher-Lee relation~\cite{Fisher1981a}. The
above equation also holds for the current through the right
contact owing to charge conservation. The Green's function
\begin{equation}
  \label{eq:G_retarded(E)}
  G^{(n)}(\epsilon) = \sum_{\alpha,n'} \frac{|u_{\alpha,n'+n}\rangle\langle
    u_{\alpha,n'}^{+}|}{\epsilon-(\epsilon_{\alpha}+n'\hbar\Omega-\I\hbar\gamma_{\alpha})}
\end{equation}
is the $t$-averaged Fourier transformed of the propagator
\eqref{eq:G_retarded(t)}. Physically, it describes the propagation
of a transmitted electron with initial energy $\epsilon$ from one
lead to the other lead undergoing scattering events with emission
($n<0$) or absorption ($n>0$) of $|n|$ photons, or being
transmitted elastically ($n=0$).

\subsection{High-frequency limit}
\label{sec:hf-limit}

The Floquet treatment of the present transport problem allows for
the implementation of a stationary perturbation scheme for driving
frequencies much larger than all other frequency scales of the
system \cite{Shirley1965a}.  This approach has recently been
extended to transport situations which are characterised by the
presence of leads \cite{Kohler2004a, Camalet2004a}; here we only
outline the derivation and refer the reader to
Ref.~\cite{Kohler2005a}. A particular benefit of this perturbation
scheme is the achievement of a physical understanding of the
transport processes by approximately mapping the time-dependent
problem to a static one with renormalised parameters. For the
static situation, in turn, the current is well known by
Eq.~\eqref{eq:Landauer}, where the transmission in the wide-band
limit reads
$T(\epsilon)=T_{\L\R}(\epsilon)=\Gamma_{\L}\Gamma_{\R}|G_{\L\R}(\epsilon)|^{2}$.
This looks similar to the driven case but with $n=0$ and therefore
we have in contrast to the driven system
$G_{\L\R}(\epsilon)=G_{\R\L}(\epsilon)$. For the static system
given by~\eqref{eq:H(t)} setting $V_{\ac}=0$, we obtain for the
transmission
\begin{equation}
  \label{eq:T}
  T(\epsilon) =
  \frac{\Gamma^2\Delta^2}{|(\epsilon-\I\Gamma/2)^2-\Delta^2|^2}
\end{equation}
assuming equal coupling to the leads ($\Gamma_{\ell}=\Gamma$).

The starting point of the approximation scheme is the unitary
transformation
\begin{equation}
  \label{eq:trafo}
  U_0(t) = \exp\left\{-\frac{\I e}{\hbar\Omega}
    V_{\ac} \sin(\Omega t) c_{\L}^\dag c_{\L}\right\},
\end{equation}
which is first applied to the quantum-well
Hamiltonian~\eqref{eq:Hwells(t)}. For sufficiently large driving
frequencies $\Omega \gg \Delta/\hbar$, a separation of time
scales is performed by this transformation. Thereby, fast oscillations
of the transformed Hamiltonian are neglected by averaging over a driving
period~\cite{Grossmann1992a,Grifoni1998a}. Finally, we arrive at the
effective Hamiltonian for the quantum wells
\begin{align}
  \label{eq:Heff}
  \bar H_{\eff} &=
  \frac{1}{\T}\int_{0}^{\T}\!\! \d t \,
  \left(U^{\dag}_0\, H_{\mathrm{wells}}(t)\,
    U_0 - \I\hbar\, U^{\dag}_0 \dot U_0 \right) \nonumber\\
  &= -\Delta_{\eff}(c_{\L}^\dag c_{\R}+c_{\R}^\dag c_{\L}),
\end{align}
which is of the same form as in the static case but with the
effective tunnelling matrix element $\Delta_{\eff} =
J_0(e V_{\ac}/\hbar\Omega)\Delta$~\cite{Grossmann1992a,Lehmann2003a}.
$J_0$ is the zeroth
order Bessel function of the first kind. Therefore, the resulting
effective transmission $T_{\eff}(\epsilon)$ with the substitution
$\Delta\rightarrow\Delta_{\eff}$ in Eq.~\eqref{eq:T} is
controllable via the driving parameters and even vanishes at zeros
of the Bessel function.

The transformation~\eqref{eq:trafo} also affects the lead--well
coupling. If we apply $U_{0}(t)$ also to $H_{\mathrm{contacts}}$ and
solve the Heisenberg equations for the lead and quantum-well
operators in the wide-band limit, we can eventually extract the new
fluctuation operator. For the left lead one finds
\begin{equation}
  \label{eq:eta(t)}
  \eta_{\L}(t) = -\frac{\I}{\hbar}\sum_q V^*_{\L q}\exp\left\{-
    \frac{\I}{\hbar}\epsilon_{\L q}(t-t_0)
     +\frac{\I e}{\hbar\Omega}V_{\ac}\sin(\Omega t)
    \right\} c_{\L q}(t_0),
\end{equation}
whereas $\eta_{\R}(t)$ remains unaffected. Now calculating the
correlation function and time-averaging it over one driving period to
neglect the \T-periodic contributions, the resulting expression
assumes the form~\eqref{eq:<xi(t)xi(t)>} but with the Fermi function
of the left lead replaced by the effective electron distribution
\begin{equation}
  \label{eq:feff}
  f_{\L,\eff}(\epsilon) = \sum^{\infty}_{n=-\infty}
  J^2_n\left(\frac{e V_{\ac}}{\hbar\Omega}\right) f_{\L}(\epsilon+n\hbar\Omega).
\end{equation}
The squares of the $n$th-order Bessel function of the first kind
in this expression weight those processes where an electron with
energy $\epsilon$ is transmitted from the left lead to the
double-well system under the emission ($n<0$) or absorption
($n>0$) of $|n|$ photons. The effective electron distribution
exhibits steps at the energies $\epsilon=\mu_{\L}+n\hbar\Omega$
and is constant elsewhere.

With the effective quantities $T_{\eff}(\epsilon)$ and
$f_{\L,\eff}(\epsilon)$ the driven problem is ascribed for fast
driving to a static one. Since $T_{\eff}(\epsilon)$ is sharply peaked
around $\epsilon=0$ and $f_{\R}(0)$ and
$f_{\L,\eff}(0)$ are constant for finite voltage, the current in the high-frequency
approximation results in
\begin{equation}
  \label{eq:I_hfa}
  \bar I = \frac{e\Gamma}{4\hbar}
  \frac{\Delta_{\eff}^{2}}{\Delta_{\eff}^{2}+(\Gamma/2)^{2}}
  \left[1+\sum_{|n|\le K(V)}
  J_{n}^{2}\left(\frac{e V_{\ac}}{\hbar\Omega}\right)\right]
\end{equation}
applying the effective parameters to the current formula~\eqref{eq:Landauer}.
Here $K(V)$ is a shorthand notation for the integer part of
$e|V|/2\hbar\Omega$.

In order to compare the transfer-matrix and the
tight-binding approach, we have to ensure that the same physical
situation is addressed. As a matching condition we compare the
transmission $T(\epsilon)$ in the time-independent case
($V_{\ac}=0$). The level splitting energy $2\Delta$ due to the
central tunnel barrier is extracted from the resonance peaks of the
doublet states computed within the transfer-matrix method. Solving
for $\Gamma$ in Eq.~\eqref{eq:T} with $\epsilon=0$ and $T(0)$ and
$\Delta$ taken from the previous calculation, the corresponding
lead--well coupling for the tight-binding system is determined.

\section{Coherent transport suppression}
\label{sec:results}

We now turn our
attention to the coherent control of current.
Tunnelling suppression in a closed, driven system is known for more
than a decade. For example for a driven bistable potential, tunnelling breaks down at exact
crossings of the quasi-energy spectrum, i.e.\ one observes the
so-called \textit{coherent destruction of tunnelling}~\cite{Grossmann1991a,Grossmann1991b}. Tunnelling
suppression has been studied in a number of
cases~\cite{Holthaus1992b,Wagner1995a,Creffield2002a}, but the
investigation in a transport context, i.e. in an open system where
an appropriate treatment of the leads is crucial, has received
attention only recently~\cite{Kohler2004a, Camalet2004a}.

\begin{figure}[t]
  \centering
  \includegraphics[width=7.5cm]{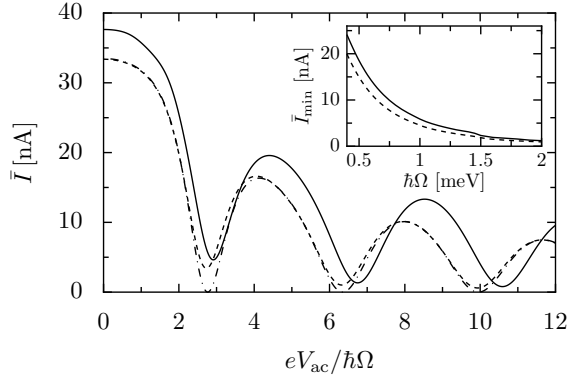}
  \caption{Average current \textit{vs.} driving amplitude obtained
  numerically from transfer-matrix (solid line) and tight-binding (dashed)
  methods. Also shown is the high-frequency
  approximation (dashed-dotted). The inset depicts the value
  of the first current minimum as a function of the driving
  frequency. Solid (transfer-matrix) and dashed
  (tight-binding) line decay approx. as $1/\Omega$. The chosen parameters
  are $\hbar\Omega=1.15\,\mathrm{meV}$, $V=6.0\,\mathrm{mV}$,
  $\Gamma=0.16\,\mathrm{meV}$ and
  $\Delta=0.23\,\mathrm{meV}$. The corresponding parameters for the barriers
  are the same as those of Fig.~\ref{fig:hetero-model}.}
  \label{fig:TMvsTB}
\end{figure}%
Surveying the time-averaged current calculated numerically from
the transfer-matrix and the tight-binding method plotted in
Fig.~\ref{fig:TMvsTB}, we observe current minima for distinct
values of $e V_{\ac}/\hbar\Omega$ for frequencies in the microwave
regime. The reason for the current suppressions becomes apparent by
comparison with the high-frequency approximation, which exhibits
minima close to those
of the transfer-matrix and tight-binding curves. The
current~\eqref{eq:I_hfa} vanishes whenever the ratio $e V_\mathrm{ac}
/\hbar\Omega$ assumes a zero of the Bessel function $J_{0}$, i.e.
for the values 2.405, 5.520, 8.654, \ldots, since then
$\Delta_{\eff}\propto J_0^2=0$. By varying the ratio between
driving amplitude and frequency, we can thus tune the tunnelling
between the two wells and thereby control the
current~\cite{Kohler2004a}. For a frequency $\Omega = 5 \Delta/\hbar$,
the analytical expression~\eqref{eq:I_hfa} shows a remarkable
agreement with the exact
tight-binding result~\eqref{eq:Ibar} for $V_{\ac} \lesssim
V$. The inset of Fig.~\ref{fig:TMvsTB} shows the minimum current
at the first suppression decays as a function of the driving frequency $\Omega$.
This is
expected from the good agreement between the numerical results and
the high-frequency approximation, because the approximation
accounts for the first order term in a perturbative scheme in
$1/\Omega$~\cite{Kohler2004a,Camalet2004a}. Higher order
contributions are included in a numerically exact calculation,
which results in a non-vanishing current at the minima. A similar
$\Omega$-dependence is observed also for the transfer-matrix
formalism.

\begin{figure}[t]
  \centering
  \includegraphics[width=7.5cm]{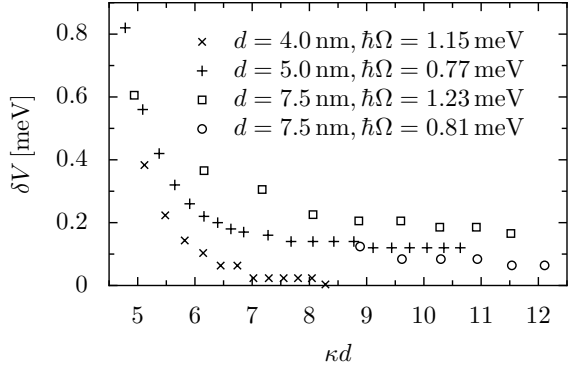}
  \caption{Deviation of the driving amplitude for the first current
  minimum from the expected first zero of $J_0$,
  $V_{0}=2.405\,\hbar\Omega$, for different barrier widths and heights. The parameters for the first three data
  sets are $d_{\C}=15\,$nm, $\bar\mu=12.0\,$meV and $V_{\C}=40\,$meV,
  whereas for the last one ($\circ$), we chose $\bar\mu=13.3\,$meV and
  $V_{\C}=80\,$meV.}
  \label{fig:WKB}
\end{figure}%
While the general shape and magnitude of the current are very
similar for both models, there still appears a small difference
in the location of the minima for the relatively low barriers chosen
in Fig.~\ref{fig:TMvsTB}. For a continuous potential, the current
assumes minima at values
of $e V_{\ac}/\hbar\Omega$ higher than those predicted by the
tight-binding description. We can understand this shift by analysing, for
given $\Omega$, the deviation $\delta V = V_{\mathrm{min}}-V_0$
of the driving amplitude $V_{\mathrm{min}}$ for which the
current exhibits its first minimum. The
amplitude $V_{0}$ corresponds to the first zero of $J_0$. In
Fig.~\ref{fig:WKB} we plot $\delta V$ as a function of $\kappa d$,
where $d=d_\L=d_\R$ and $\kappa=[2m(V-\bar{\mu})/\hbar^2]^{1/2}$, i.e.
$\kappa d$ is the instanton action in units of $\hbar$ and
$\exp(-2\kappa d)$ is the WKB transmission probability of the outer barriers in Fig.~\ref{fig:hetero-model}. Here
$V=V_{\L}=V_{\R}$ is the corresponding barrier height and
$\bar{\mu}=(\mu_{\L}+\mu_{\R})/2$ denotes the average chemical potential
representing approximately the mean energy of the resonance doublet.

If the width of the outer barriers is kept fixed, $\delta V$
decreases for growing $\kappa d$ because then the resonance energies
are further away from the barrier edge. Therefore, the wave
functions of the well states become more localised. This situation
corresponds in the tight-binding picture to a lead--well coupling
$\Gamma$ that is almost energy independent and thus reproduces
the wide-band limit. Furthermore, this argument is used to explain
the smaller deviation observed with thinner barriers for the same
$\kappa d$, since $V_{\L}$ is much larger in that case.

As can be seen by comparing data sets for different central
barrier heights in Fig.~\ref{fig:WKB}, an increase of the height of the central
barrier $V_{\C}$
reduces the level splitting $2\Delta$, that is, the overlap
between the localised states in the left and right well in a
tight-binding description. Thus the tight-binding
and the transfer-matrix results converge as a function of the barrier height. Finally, it is
important to note that varying any of the barriers affects the
transmission properties of the whole heterostructure, in contrast
to the tight-binding model, where the different coupling
parameters can be controlled independently.

\section{Conclusions}
\label{sec:conlusio}

We have demonstrated that the dc current across a double-well
heterostructure can be suppressed by the purely coherent influence of an oscillating gate voltage. We have used a transfer-matrix method as an
exact approach to compute tunnelling currents through such a
system. We compared these results to that obtained from
a tight-binding Floquet description. In particular, we find that the current
suppression is controlled by the ratio of the driving frequency and amplitude.
This can be understood by exploring the high-frequency limit
within the tight-binding formalism. In this perturbative scheme,
the time-dependent system is mapped onto a static one with
renormalised parameters, that is, with an {effective hopping
matrix element} accounting for inter-well coupling and with an
{effective electron distribution} for the attached left lead.
Since the effective inter-well coupling depends on the ratio
between driving amplitude and frequency, the transport properties
of the double well can be adjusted through the driving parameters,
with the effective behaviour ranging from transport through an
almost open channel to a regime of rare tunnel events.

The results presented in this work strongly support the idea that
transfer-matrix and tight-binding descriptions of quantum
transport are equivalent provided the barriers are sufficiently high.
In this case the lowest resonance states in the wells are rather
localised and consequently the tight-binding description becomes
accurate. For a proper choice of parameters, we find a good
agreement between the exact transfer-matrix calculation and the
results obtained within the tight-binding formalism. The study
presented here shows that the coherent control of time-dependent
electron transport can be investigated with current semiconductor
nanotechnology.

\section*{Acknowledgements}

We thank S\'ebastien Camalet, Gert-Ludwig Ingold, and J\"org Lehmann
for helpful discussions. Two of us (M.R. and F.S.) wish to thank
Mathias Wagner for introducing us to a numerical code based on Ref.
\cite{Wagner1995a}. This work has been supported through a joint
Germany-Spain Acci\'on Integrada (no. HA2003-0091). Financial
support is also acknowledged from MEC (Spain), Grant no.
BFM2001-0172, Fundaci\'on Ram\'on Areces, and DFG (Germany),
Graduiertenkolleg~283 and Sonderforschungsbereich~486. We also thank
the Centro de Computaci\'on Cient{\chardef\i="10 \'\i}fica (UAM) for
technical and computational support.

\end{document}